\documentclass{article}

\usepackage{algorithm}
\usepackage{algorithmic}
\usepackage{bnaic}
\usepackage{graphicx}
\usepackage{tikz}
\usepackage{amsmath}
\usepackage{amsthm}
\usepackage[pdfusetitle]{hyperref}

\theoremstyle{definition}
\newtheorem{definition}{Definition}[section]

\title{\textbf{\huge The Complexity of Rummikub Problems}} 
\author{Jan N. van Rijn \and Frank W. Takes \and Jonathan K. Vis}
\date{\textit{Leiden Institute of Advanced Computer Science, Leiden University, The Netherlands}}

\hypersetup{pdftitle={The Complexity of Rummikub Problems},pdfauthor={Jan N. van Rijn, Frank W. Takes, Jonathan K. Vis}}

\begin{document}


\ttl
\thispagestyle{empty}

\begin{abstract} 
\noindent
Rummikub is a tile-based game in which each player starts with a hand of $14$ tiles.
A tile has a value and a suit. 
The players form sets consisting of tiles with the same suit and consecutive values (runs) or tiles with the same value and different suits (groups).
The corresponding optimization problem is, given a hand of tiles, to form valid sets such that the score (sum of tile values) is maximized. 
We first 
present an algorithm that solves this problem in polynomial time. 
Next, we analyze the impact on the computational complexity when we generalize over various input parameters. 
Finally, we 
attempt to better understand some aspects involved in human play by means of an experiment that considers counting problems related to the number of possible immediately winning hands.  
\end{abstract}



\section{Introduction}
\label{sec:introduction}


Games are generally played in a precisely defined and contained environment with fixed rules and behavior. This allows humans to witness and easily interpret the power and limitations of algorithms for solving computational problems~\cite{Herik2002}. 
In this paper we consider the complexity for the game of Rummikub, a tile game introduced in the 1930s  
and well-known in, amongst others, northwestern Europe. 
The game is played in the following way. 

Rummikub uses a set of tiles containing $k > 0$~\emph{suits} (represented by colors). 
For each suit we have $n > 0$ tiles, numbered from $1$ to $n$. 
The number of a tile represents the $\mathit{value}$ of that tile. 
In the game of Rummikub $m > 0$ copies of the $n \times k$ tiles are present in the full \emph{tile set}.
Additionally, a small number of $j \geq 0$ \emph{jokers} are allowed to represent any other tile. 
In the original game of Rummikub $m = 2$ copies are used, both consisting of $k = 4$ suits that each contain $n = 13$~tiles. 
Generally, $j = 2$~jokers are used, summing to $106$~tiles.
Indeed, the game can also be played using two card decks, and we use the card suit symbols for convenient tile visualization of the original game.

\begin{figure}[t!]
	\begin{center}
		\begin{tikzpicture}

\draw[rounded corners=1mm] (0, 0.2) rectangle (0.8, 1.4);
\node[draw, circle, text=blue] at (0.4, 1.0) {\textbf 6};
\node[text=blue] at (0.62, 0.4) {$\boldsymbol\clubsuit$};

\draw[rounded corners=1mm] (1, 0.2) rectangle (1.8, 1.4);
\node[draw, circle, text=blue] at (1.4, 1.0) {\textbf 7};
\node[text=blue] at (1.62, 0.4) {$\boldsymbol\clubsuit$};

\draw[rounded corners=1mm] (2, 0.2) rectangle (2.8, 1.4);
\node[draw, circle, text=blue] at (2.4, 1.0) {\textbf 8};
\node[text=blue] at (2.62, 0.4) {$\boldsymbol\clubsuit$};

\draw[rounded corners=1mm] (3, 0.2) rectangle (3.8, 1.4);
\node[draw, circle, text=blue] at (3.4, 1.0) {\textbf 9};
\node[text=blue] at (3.62, 0.4) {$\boldsymbol\clubsuit$};

\draw[rounded corners=1mm] (5, 0.2) rectangle (5.8, 1.4);
\node[draw, circle, text=blue] at (5.4, 1.0) {\textbf 8};
\node[text=blue] at (5.62, 0.4) {$\boldsymbol\clubsuit$};

\draw[rounded corners=1mm] (6, 0.2) rectangle (6.8, 1.4);
\node[draw, circle, text=blue] at (6.4, 1.0) {\textbf 9};
\node[text=blue] at (6.62, 0.4) {$\boldsymbol\clubsuit$};

\draw[rounded corners=1mm] (7, 0.2) rectangle (7.8, 1.4);
\node[draw, circle, inner sep=2pt, text=blue] at (7.4, 1.0) {\textbf{10}};
\node[text=blue] at (7.62, 0.4) {$\boldsymbol\clubsuit$};

\draw[rounded corners=1mm] (0, 2) rectangle (0.8, 3.2);
\node[draw, circle, text=red] at (0.4, 2.8) {\textbf 3};
\node[text=red] at (0.62, 2.2) {$\boldsymbol\diamondsuit$};

\draw[rounded corners=1mm] (1, 2) rectangle (1.8, 3.2);
\node[draw, circle, text=blue] at (1.4, 2.8) {\textbf 3};
\node[text=blue] at (1.62, 2.2) {$\boldsymbol\clubsuit$};

\draw[rounded corners=1mm] (2, 2) rectangle (2.8, 3.2);
\node[draw, circle] at (2.4, 2.8) {\textbf 3};
\node at (2.62, 2.2) {$\boldsymbol\heartsuit$};

\draw[rounded corners=1mm] (4, 2) rectangle (4.8, 3.2);
\node[draw, circle, text=red] at (4.4, 2.8) {\textbf 3};
\node[text=red] at (4.62, 2.2) {$\boldsymbol\diamondsuit$};

\draw[rounded corners=1mm] (5, 2) rectangle (5.8, 3.2);
\node[draw, circle, text=blue] at (5.4, 2.8) {\textbf 3};
\node[text=blue] at (5.62, 2.2) {$\boldsymbol\clubsuit$};

\draw[rounded corners=1mm] (6, 2) rectangle (6.8, 3.2);
\node[draw, circle] at (6.4, 2.8) {\textbf 3};
\node at (6.62, 2.2) {$\boldsymbol\heartsuit$};

\draw[rounded corners=1mm] (7, 2) rectangle (7.8, 3.2);
\node[draw, circle, text=green!80!black] at (7.4, 2.8) {\textbf 3};
\node[text=green!80!black] at (7.62, 2.2) {$\boldsymbol\spadesuit$};

\end{tikzpicture}
	\end{center}
	\vspace{-4mm}
	\caption{Two valid groups and two valid runs.} 
	\label{fig:rummikub}
\end{figure}
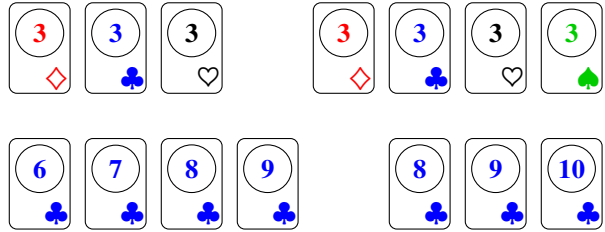

Rummikub is typically played with two to four players. 
The \emph{pool} is defined to be the stack of all tiles in a random order. 
Initially, each player is provided with a \emph{hand} consisting of $14$ tiles from the pool. 
Then, players take turns in playing tiles. 
A tile that is \emph{played} moves from the player's closed hand to the open \emph{table}.  
Each turn, a player can play as many tiles as he wants. 
If a player chooses not to play any tiles, the player takes a tile from the pool, if it still contains any. 
Any tile that is played must be an element of either a group or a run. 
A \emph{group} is defined as a set of at least three tiles of the same value but a different suit. 
A \emph{run} is as a set of at least three tiles of the same suit but with consecutive values. 
Examples of groups and runs are given in Figure~\ref{fig:rummikub}. 
Players are also allowed to \emph{rearrange} tiles that are on the table as long as this again finally results into (possibly completely different) valid runs and groups. 
At the end of a turn all tiles that were originally on the table must still be on the table, referred to as the \emph{table constraint}. 

At the start of the game all players are considered to be in \emph{quarantine}.
A player in quarantine is not allowed to play tiles until he forms a number runs and groups whose total value exceeds a certain threshold $\theta$ (typically $\theta = 30$). This is sometimes referred to as the \emph{initial meld}. As a consequence, players in quarantine cannot play tiles as part of runs or groups that are already on the table. 
After playing tiles of which the sum exceeds the threshold value, the player is no longer in quarantine. 
The first player who is able to play all his tiles wins and ends the game.
Alternatively, the game ends whenever the pool is empty and all players choose not to play any tiles. 
Then some scoring function is applied based on the sum of tile values, either remaining in the hand or played on the table, determining the winner. 

The game of Rummikub contains an optimization problem, which we call the \emph{Rummikub puzzle}. 
Given the combination of a hand and a table, the goal is to maximize the number of points that can be obtained by making valid runs and groups. 
The main contribution of this paper is a classification of which generalizations of the Rummikub puzzle are computationally difficult, and which ones are not. 
Specifically, we give an algorithm that solves the optimization problem in polynomial time, and discuss how different alternative input parameters influence the complexity of that algorithm. 
Subsequently, we use the algorithm to address a counting problem regarding the percentage of hands that can be played in one move in an attempt to gain more insight in possible strategies for the multi-player game. 

The remainder of this paper is organized as follows. 
First, in Section~\ref{sec:relatedwork} relevant previous work on Rummikub and similar games is discussed. 
Next, a formal problem definition and characterization of the problem's input parameters is given in Section~\ref{sec:theory}. 
In Section~\ref{sec:polyalgo} a polynomial algorithm for a particular class of Rummikub puzzles is described. 
Section~\ref{sec:counting} uses this algorithm to address the aforementioned counting problem.
Finally, Section~\ref{sec:conclusion} provides concluding remarks and suggestions for future work. 


\section{Related work}
\label{sec:relatedwork}

Rummikub is somewhat related to so-called ``hand-making games''. 
In~\cite{Iwama2002} an online model has been developed that attempts to solve the ``removable online knapsack problem'', a version of the knapsack problem where items are handled in a certain order and the player needs to decide whether such an item will be added to the knapsack or not. 
Items that are added can later be removed from the knapsack, but items that are neglected or removed cannot be added later on. 
The authors claim that this model can be applied to hand-making games such as Rummy. 
A simplified version of the game Gin Rummy is used to compare various self-play methods against each other in~\cite{Kotnik2003}. 

The Rummikub puzzle as it is considered in this paper, has not been studied much. 
In~\cite{Rijn2012} an unsuccessful attempt was made to prove a generalized version of the Rummikub puzzle NP-complete. 
An integer linear programming method for solving the Rummikub puzzle is proposed in \cite{Hertog2006}. 
There, it is stated that the optimization problem of obtaining as many points in one move as possible is ``very difficult, since the number of possible combinations are enormous''. 
Integer linear programming is an optimization technique often used to approximate intractable problems. 
As we will show, this optimization problem can actually be solved in polynomial time. 


\section{Preliminaries}
\label{sec:theory}

The results presented in this work focus on the Rummikub puzzle:

\begin{definition}{\emph{Rummikub puzzle}}\label{def:rummikubpuzzle} \\
Given a subset of the Rummikub tile set of $n \times k \times m$ tiles with $n$ values, $k$ suits and $m$ copies of each tile, 
form valid sets of runs and groups such that the score (sum of used tile values) is maximized. 
\end{definition} 

\noindent
Note that this optimization problem is the equivalent of minimizing the value of the unused tiles in the subset. 
The problem is closely related to the following decision problem: 
given a subset of the Rummikub tile set, can we use all tiles in valid runs and groups? 
This decision problem can trivially be reduced to the optimization problem, as the outcome of the decision problem is true if and only if the solution to the optimization problem is equal to the sum of the values of the tiles in the particular subset. 

This optimization problem is related to the original multi-player game. Each turn a player uses the tiles in his hand combined with the tiles on the table to form valid groups and runs. 
The particular subset for which the Rummikub puzzle is solved, is the union of the tiles on the table and the tiles in the player's hand.
The aforementioned table constraint is easily satisfied by disregarding solutions in which unused tiles were on the table. 
The availability of a constant number of jokers can also be accommodated, as we discuss in Section~\ref{sec:algogame}.


To reason about complexity problems, we should also define which variables are parameters of the problem. %
Here we distinguish between tile set parameters and the set size parameter. 
The \emph{tile set parameters} are the previously mentioned $k$ for the number of suits, $n$ for the number of tiles of each suit and $m$ for the number of copies of $k \times n$ tiles in the full tile set.
The \emph{set size parameter} $s$ is the minimal size of a set (either a group or a run), which is equal to $s = 3$ in the original Rummikub game.

The question of defining the input size of the problem is open to some interpretation with respect to which parameters should be considered. 
For example, one could say that by increasing the number of suits $k$ we should also increase the minimal set size, for example by fixing it at $s = k - 1$. 
This would also mean increasing the minimal run length, which is not always possible if $k > n$. 
To avoid all kinds of ambiguous situations that are undefined in the original game (such as replacing the minimal set size parameter by separate parameters for the minimum run length and minimal group size), we only consider tile set variables $n$, $k$ and $m$ as input parameters, and fix the set size parameter at $s = 3$.


\section{Polynomial algorithm} 
\label{sec:polyalgo}

Given a particular subset of the Rummikub tile set, we aim for an algorithm that makes valid sets of runs and groups and thus 
solves the Rummikub puzzle optimization problem posed in Definition~\ref{def:rummikubpuzzle}. 
For a given hand, for each tile in this hand we can choose to use it in a run, use it in a group or not to use it in either one. 
Indeed, given a hand of $t$ tiles, a brute-force approach would find the solution after exploring $3^t$ states. 
Clearly, this does not scale and is not feasible for the original Rummikub game. 

In this section we give a polynomial algorithm for solving the Rummikub puzzle. 
The section starts by describing group formation in Section~\ref{sec:forminggroups} and a representation of the state space in Section~\ref{sec:states}.
The proposed algorithm for exploring this state space is then the subject of Section~\ref{sec:algo}. 
The complexity is analyzed in Section~\ref{sec:complexity} and in Section~\ref{sec:algogame} we discuss various smaller aspects and constraints that need to be satisfied in Rummikub. 

\subsection{Forming groups}
\label{sec:forminggroups}

We consider the task of forming groups consisting of tiles of the same value but a different suit given $k$ suits and $m$ tile set copies. 
We are interested in the number of ways $G(k, m)$ in which this can be done. 

For the variant of Rummikub where $k = 4$ and $m = 1$ we can choose to either not form any groups, choose either one of the four ways to form groups of size three or choose to form one group of size four, summing to six possible ways of forming groups, denoted as $G(4, 1) = 1 + 1 + 4 = 6$. 
Note that the alternative way of generalizing over $k$ with $s = k - 1$ as discussed in Section~\ref{sec:theory} would make $G(k, 1)$ linear in $k$: 
$G(k, 1) = 1 + {{k}\choose{k-1}} + {{k}\choose{k}} = 1 + k + 1 = k + 2$. 
To generalize over $k$ (but still with constant $m = 1$) given our definition of a constant $s$, the number of groups is one (no groups) plus the number of ways we can form groups of size $s$ up to size $k$:	

$$G(k, 1) = 1 + \sum_{i=s}^{k} { {k}\choose{i} } $$

\noindent
To generalize over $m$ we first observe that for $m = 2$ we have a copy of the full tile set with which we can again form groups in $G(k, 1)$ possible ways. 
Indeed, a trivial upper bound for $G(k, m)$ is then: 

$$G(k, m) \leq G(k, 1)^m$$

\noindent
However, this upper bound is not necessarily tight: there are combinations that are counted more than once, for example because the order in which we form groups does not matter and because groups may have overlapping tiles. 
For example, for $m = 2$ and $k = 5$ selecting the two groups $\{\clubsuit, \heartsuit, \spadesuit\}$ and $\{\heartsuit, \spadesuit, \diamondsuit, \triangle\}$ achieves the same tile usage as selecting groups $\{\heartsuit, \spadesuit, \diamondsuit\}$ and $\{\clubsuit, \heartsuit, \spadesuit, \triangle\}$. 
Although the upper bound presented here is not tight, we expect that the number of ways to form groups is still exponential or perhaps even factorial. 
Thus, considering all possible ways of forming groups (and using the remaining tiles in runs) is not a good idea for an algorithm for solving Rummikub. 
In the next section we investigate the other way around: first making runs, and then using the remaining tiles in groups. 

\subsection{State space}
\label{sec:states}

We propose a dynamic programming approach in which the \emph{state} of the puzzle is denoted by a combination of (1) the current value $v$ (with $1 \leq v \leq n$) that we are considering and (2) the current length of runs of each of the $m$ copies of each suit $k$, which we refer to as the current $\mathit{runs}$.
An important thing to note is that given $s = 3$, only runs of length $0$, $1$, $2$ and $3$ or larger are to be distinguished, so  $s + 1 = 4$ different cases (and, importantly, not $n$ different cases). 

Consider $m = 1$. 
At each tile value for each suit we can choose to either continue a run, increasing the length of the run by one, or to stop the run (either because we use the tile in a group, or because we have no tile to continue the run), resetting the length of that run to zero. 
In case a run increases from length two to three at value $v$, it becomes \emph{valid} and the obtained score is the sum of the run so far, i.e., $(v-2)+(v-1)+v$. 
The algorithm also scores at each tile value the score obtained by either making groups of that value or by continuing valid runs using a tile of the current value $v$. 
In such a way, a run of length three does not have to be distinguished from a run of length larger than three, as adding a tile to the run will simple give the value of that tile as added score. 
If an existing run is not continued, the run ends, its length is reset to zero and no score is obtained for this run. 

For $m = 2$, the number of different configurations of $\mathit{runs}$ for a particular suit is $10$: $\{0, 0\}$, $\{0, 1\}$, $\{0, 2\}$, $\{0, 3\}$, $\{1, 1\}$, $\{1, 2\}$, $\{1, 3\}$, $\{2, 2\}$, $\{2, 3\},$ and $\{3, 3\}$. 
For general $m$ and $s$ the number of configurations $f(m)$ is given by the number of distinct permutations of size $m$ in a multiset of $k$ distinct elements of multiplicity $m$, denoted by the multinomial coefficient:

$$f(m, s) = {{s + 1} \choose m}_m$$ 

\noindent
While this may not seem polynomial, for the Rummikub set size parameter $s = 3$, we get exactly the tetrahedral (or triangular pyramidal) numbers: 

$$f(m) = {4 \choose m}_m = \frac{(m+1) \cdot (m+2) \cdot (m+3)}{6}$$ 

\noindent
So the size of the considered state space is at most $n \times k \times f(m)$, i.e., it is polynomial in $n$, $k$, and $m$.

\subsection{Algorithm}
\label{sec:algo} 

\begin{algorithm}[t]                       
	\caption{\textsc{maxScore}}  
	\label{alg:polalgo}          
	                
	\begin{algorithmic}[1]
		\STATE \textbf{input: } $\mathit{value}$, $\mathit{runs}$[$k\times f(m)$] \hfill (the state)
		\STATE \textbf{output: } maximum score \hfill (given the input state)
		
		\vspace{2mm}
		\IF {$\mathit{value} > n$}
			\STATE \textbf{return} $0$
		\ENDIF		
		\IF {$\mathit{score}$[$\mathit{value}$, $\mathit{runs}$] $ > -\infty$}
			\STATE \textbf{return} $\mathit{score}$[$\mathit{value}$, $\mathit{runs}$]
		\ENDIF		
		\vspace{2mm}		
		\FOR {$\mathit{runs'}$, $\mathit{runscores}$ $\in$ \textsc{makeRuns}$($$\mathit{runs}$$)$} 	
				\STATE $\mathit{groupscores} \leftarrow $ \textsc{totalGroupSize}$(\mathit{hand} \setminus \mathit{runs'}) \cdot \mathit{value}$ 
				\STATE $\mathit{result} \leftarrow$ $\mathit{groupscores} + \mathit{runscores} + \textsc{maxScore}$($\mathit{value} + 1$, $\mathit{runs'})$
				\STATE $\mathit{score}[\mathit{value}$, $\mathit{runs}$] $\leftarrow \max(\mathit{result}, \mathit{score}$[$\mathit{value}$, $\mathit{runs}$]$)$
		\ENDFOR	
		\vspace{2mm}
		\STATE \textbf{return} $\mathit{score}$[$\mathit{value}$, $\mathit{runs}$] 
	\end{algorithmic}
\end{algorithm}

The proposed recursive procedure to compute the maximum score given some hand of Rummikub tiles is outlined in Algorithm~\ref{alg:polalgo}. 
The algorithm represents the entire state space using an $n \times k \times f(m)$ multi-dimensional array named $\mathit{score}$ which contains the maximum score that can be obtained given this state of the puzzle.
This array is initialized to $-\infty$. 
The global variable $\mathit{hand}$ contains the particular subset of tiles for which we are running the algorithm, and is implicitly used at various moments. 
The algorithm is initially called with $\mathit{value}$ $= 1$ and with a $\mathit{runs}$ vector of size $k$ initialized for each suit to a multi-set of size $m$ with precisely $m$ zeros. 

So, starting at tile value $1$, the algorithm determines how the current $\mathit{runs}$ can be extended given the tiles that we have available of the current value (line~9). 
Then for each possible extension of the $\mathit{runs}$ (determined by \textsc{makeRuns}$($...$)$, further discussed in Section~\ref{sec:complexity}) it computes the score obtained when using the remaining tiles ($\mathit{hand} \setminus \mathit{runs'}$) in as large as possible groups, determined on line~10 by \textsc{totalGroupSize}$($...$)$. 
For each possible $\mathit{runs'}$ vector, a recursive call for $\mathit{value} + 1$ is made, of which the resulting score is added to the score of the groups and runs made at the current value (line~11). 
The $\mathit{score}$ array is then updated with this total score (line~12). 
Finally, the algorithm returns the highest possible score given the current $\mathit{value}$ and $\mathit{runs}$ (line~13). 
Upon a later query for the maximum score given the current $\mathit{value}$ and $\mathit{runs}$, the previously computed score can be returned (line~7), i.e., the dynamic programming aspect. 


Note that during the execution of the algorithm it is easy to keep track of the solution by simply storing for each state $\mathit{score}$[$\mathit{value}$, $\mathit{runs}$] which following state $\mathit{score}$[$\mathit{value} + 1, \mathit{runs}$] yielded its score.
Furthermore, we mention that many smaller optimizations can be made, for example by ending the recursion when a score equal to the sum of all values of the tiles in the tile set (perhaps even subtracting the value of tiles that can never be part of any run or group) is reached. 

\subsection{Complexity}
\label{sec:complexity}

The \textsc{makeRuns}$($...$)$ function on line~9 of Algorithm~\ref{alg:polalgo} iterates over all possible ways of making runs given the current configuration of $\mathit{runs}$. 
In Section~\ref{sec:states} we have shown that the number of these configurations for some suit is bounded by $f(m)$, showing that the size of the state space is bounded by a polynomial. 
Now, to derive the time complexity of the proposed algorithm, let us consider the number of possibly ways of continuing runs, i.e., the number of states generated by the \textsc{makeRuns}$($...$)$ function. 


Continuing runs for $m = 1$ is trivial; we simply continue a run of a particular suit if we have the correct tile to do so, or we do not and instead use the tile in a group. 
For $m = 2$, for a particular suit we may continue none of the runs, both of the runs, or just one of the runs (assuming we have sufficient tiles of that suit and value to do so). 
In the latter case, if the length of these runs differs, both of the runs have to be tried, as for a particular suit, given a run of length $1$ with tile $\langle8\rangle$ and a run of length $3$ with tiles $\langle6, 7, 8\rangle$, if we are currently considering value $9$ and have one tile with this value, then we cannot decide to which run the $9$ should be added without trying both. 

We note that a sequence of $\mathit{runs}$ for a particular suit is always of the form $\{0, 0, 0, 1, 1, 1, 1, 2$, $3, 3\}$ (in this case for $m = 10$), more briefly written as a multiset $\{0^3, 1^4, 2^1, 3^2\}$ or even shorter as $\langle3, 4, 1, 2\rangle$. 
Thus, in this example we have two runs that are complete, one (incomplete) run of two tiles, four (incomplete) runs of one tile and three runs that do not contain any tiles.
Important to note is again the fact that using only four integer values, we can encode the $\mathit{runs}$ for any $m$. 
If we have $x$ tiles (with $x \leq m$) of some suit in our hand, we consider using $x' \in \{0, 1, \ldots, x\}$ tiles ($x + 1$ options) to complete the current runs. 
This can again be done in ${4 \choose x'}_{x'}$
 ways, which, as we discussed in Section~\ref{sec:states}, is equal to 
 $(x'+1) \cdot (x'+2) \cdot (x'+3) / 6$. 
As $x$ is at most equal to $m$, the number of ways in which we can continue runs of a particular suit configuration is again polynomial in $m$. 
For each of the $k$ suits, every continuation of runs has to be tried, so the number of ways to continue a run is bounded by $((m+1) \cdot f(m))^k$, which is $O((m^4)^k$.
However, most importantly, we note that the number of states generated by \textsc{makeRuns}$($...$)$ is polynomial in $m$, i.e., the number of suits $k$ is the only parameter over which we cannot generalize in polynomial time. 
The above leads us to the following conclusions about the complexity of Rummikub: 

\begin{enumerate}
	\item The traditional Rummikub puzzle with fixed $m$ and $k$ can be solved in polynomial time. In fact, the presented algorithm solves the puzzle in linear $O(n)$ time, which is optimal. 
	\item The state space of the Rummikub puzzle is at most of size $O(n \cdot k \cdot m^4)$, so polynomial in all input parameters $n$, $m$ and $k$.
	\item The Rummikub puzzle can be solved in polynomial time $O(n \cdot m^4)$ when the number of tile values $n$ and the number of tile set copies $m$ are considered as input parameters. 
	\item Our algorithm for solving Rummikub runs in exponential $O(n \cdot (m^4)^k)$ time when in addition to $n$ and $m$ the number suits $k$ is considered as input parameter. 
\end{enumerate}

\subsection{Discussion}
\label{sec:algogame}

The proposed algorithm for solving Rummikub puzzles can be employed in the game of Rummikub as discussed in Section~\ref{sec:introduction}. 
In each turn we run Algorithm~\ref{alg:polalgo} on the union of all tiles in the player's hand and all tiles on the table. 
Then we play the move that results in the highest score, of course taking the initial meld at the beginning of the game into account. 

Furthermore, the table constraint can easily be satisfied by checking at each value whether or not all tiles of a formed group are actually used in the newly formed hand. 
For newly started runs, it may not be clear whether or not a tile of value $v-2$ is actually used in a run until we are at value $v$ (when a valid run is formed, or not). 
This can be solved by means of bookkeeping: we use a constant size sliding vector of two counters indicating how many tiles of each suit we used at iteration $v-1$ and $v-2$, and disregard the current solution should we find that we ended up disregarding a run of which the tiles were previously on the table, which would have meant violating the table constraint.

The availability of a constant number of jokers can also be
accommodated in the puzzle variant of Rummikub. Usually, there are two
joker tiles which can both be used to represent any regular tile in
the game. Independent of whether these tiles are actually present,
i.e., there can be configurations where a joker represents a value of a certain suit
while the two copies of this value and suit are also present. Without this
rule the addition of jokers would be trivial. For
Algorithm~\ref{alg:polalgo} this means that we can add a dimension to
the search space representing the number of jokers that we used so
far. Then for each value in $\{1,2,\ldots,n\}$ we substitute a joker
for each missing tile of this value (provided there are enough jokers
remaining). For $j = 2$ we increase the search space with a
factor~$3$, i.e., $0$, $1$, or $2$~jokers remaining.

To accommodate the possibility of having more than two values of a
particular suit we can still use the same construction. The method of
creating groups must be adjusted because potentially more groups can be created.
We also have more possibilities of having both groups and runs to consider. 
The search space will not increase beyond the aforementioned constant factor, keeping
Algorithm~\ref{alg:polalgo} for the original Rummikub puzzle in $O(n)$.

Jokers do not contribute to the score of the configuration. However,
$25$~penalty points per joker are given in rare occasions where a joker
cannot be used in a configuration. Note that this minor score modification still
allows for the conversion of the optimization problem to its
corresponding decision problem.


\section{Counting winning hands} 
\label{sec:counting}

Now that we have better idea of the theoretical complexity of Rummikub, let us consider an aspect of the game that has some importance when Rummikub is played by a human. 
We study the number of hands that can be won in one move, which has some relevance during the game, as it provides insight in the likelihood of the game ending at different points in time. 

Counting problems have attracted some attention in the literature. 
For many games, the size of the state space has either been estimated or exactly determined~\cite{Herik2002}. 
For one-player games (puzzles), different counting problems are typically
addressed. 
For example in \cite{Yan2004}, the number of Patience games that cannot be won 
is estimated. 
In this section we address a similar question: 
Given a hand of $t$ tiles, what percentage of all possible hands of this size can be played in one move? 
We refer to such hands as \emph{winning hands}.

The number of possible hands of a given size can be determined using the
multinomial coefficient: 
$$h(n,m,k,t) = {n \cdot k \choose t}_m$$

\noindent Effectively, the problem is reduced to enumerating over all possible hands of this size 
and then using Algorithm~\ref{alg:polalgo} to determine whether we can use all tiles in valid groups and runs. 
As the multinomial numbers grow exponentially, this approach becomes already impractical for rather small input. 

A more practical method to count the number of winning hands is the following. 
We start by generating all partitions of a hand of size $t$. 
In fact, we only consider partitions consisting of subsets of size $3$, $4$ and $5$, so that each partition potentially represents a valid run or group. 
Note that we do not need to consider partitions of size larger than five, as every number larger than three can be written as a summand of $3$, $4$ and $5$.
Depending on how this method enumerates the partitions, some winning hands can 
be encountered multiple times. For this reason, we also need to store all winning
hands in memory. This way we can check whether a hand was already counted and avoid counting duplicates.

\begin{table}[t!]
\begin{center}
\caption{
  Number of sets of tiles of a given size with $n = 13$, $m = 2$ and $k = 4$. 
}
\normalsize 
\label{tab:count}
\begin{tabular}{r r r r}
\hline
$t$ & hands of size $t$ & winning hands of size $t$ & ratio ($\cdot 10^{-7}$) \\
\hline
\small
  $14$ &             $37,\!418,\!772,\!170,\!780$ &        $10,\!232,\!524$ & $2.73$ \\
  $15$ &            $148,\!416,\!376,\!650,\!360$ &        $75,\!493,\!324$ & $5.09$ \\
  $16$ &            $553,\!693,\!464,\!464,\!595$ &       $167,\!019,\!567$ & $3.02$ \\
  $17$ &        $1,\!949,\!530,\!720,\!153,\!380$ &       $266,\!275,\!320$ & $1.37$ \\
  $18$ &        $6,\!497,\!700,\!004,\!347,\!370$ &   $1,\!285,\!155,\!978$ & $1.98$ \\
  $19$ &       $20,\!554,\!261,\!726,\!376,\!560$ &   $3,\!043,\!378,\!964$ & $1.48$ \\
  $20$ &       $61,\!854,\!641,\!867,\!215,\!015$ &   $5,\!281,\!155,\!009$ & $0.85$ \\
  $21$ &      $177,\!450,\!513,\!642,\!518,\!480$ &  $18,\!897,\!450,\!032$ & $1.06$ \\
  $22$ &      $486,\!216,\!174,\!534,\!733,\!370$ &  $45,\!490,\!938,\!770$ & $0.94$ \\
  $23$ &  $1,\!274,\!559,\!907,\!320,\!479,\!780$ &  $83,\!353,\!290,\!572$ & $0.65$ \\
  $24$ &  $3,\!201,\!331,\!817,\!672,\!585,\!415$ & $241,\!746,\!095,\!133$ & $0.76$ \\
  $25$ &  $7,\!715,\!065,\!735,\!511,\!650,\!152$ & $570,\!816,\!408,\!020$ & $0.74$ \\
  $26$ & $17,\!862,\!050,\!779,\!716,\!207,\!204$ & $1,\!076,\!455,\!604,\!342$ & $0.60$ \\
\hline
\end{tabular}
\end{center}
\end{table}

Table~\ref{tab:count} shows for each hand size $t$ the total number of such hands and the number of winning hands, whereas the column ``ratio'' denotes the ratio of hands that are winning. 
From this table it follows immediately that at the start of a game, there is a chance of $0.0000273\%$ that the starting player will win the game in one move.
As the size of the hands is relatively small compared to the number of tiles in a
game of Rummikub, 
it seems plausible that the ratio of winning hands is rather low.
\begin{figure}[!b]
  \begin{center}
    \includegraphics[angle=270,scale=.52]{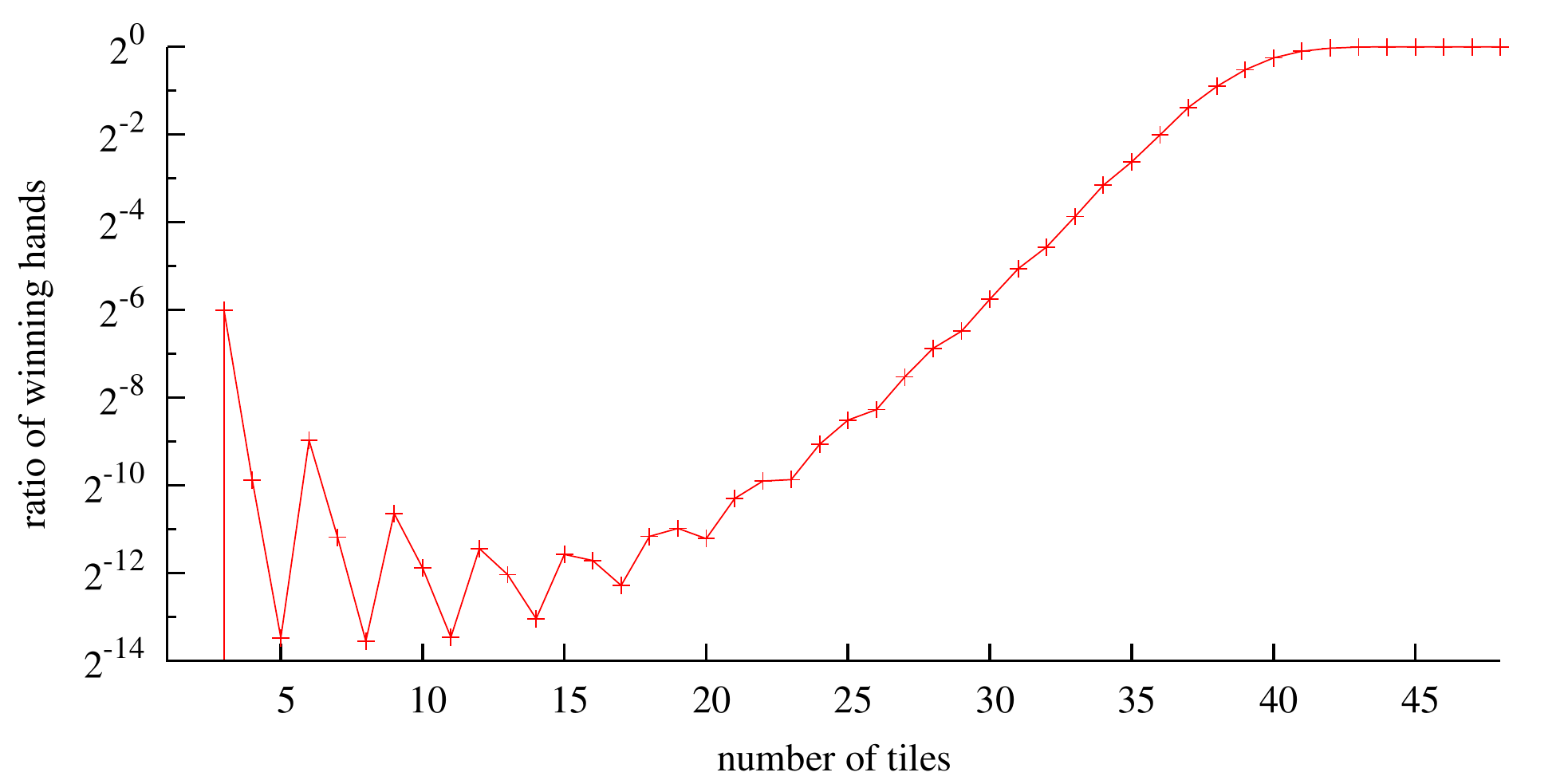}
    \caption{The ratio of winning hands of a given size for
             Rummikub with $n = 6$, $m = 2$ and $k = 4$.}
	 \label{fig:counts}
  \end{center}
\end{figure}
One observation from Table~\ref{tab:count} is that having more tiles does not necessarily 
improve the percentage of the hands that can be played in one move.
However, 
it is reasonable to assume that at some point the ratio of winning hands grows considerably. 

It can be seen that when a hand consists of all tiles in the game it is by 
definition a winning hand, as there are trivial ways to play all tiles. 
The same holds when a hand consists of all tiles except one or two; 
regardless of which tiles are missing, we can play all tiles as groups. 
Also for hands consisting of all tiles except three or four it holds that it is
always a winning hand; this can be verified using Algorithm~\ref{alg:polalgo}. 
Hands consisting of fewer tiles are not necessarily winning hands. 
Another observation from Table~\ref{tab:count} is that overall, the ratio seems to
decrease as the number of tiles grows, but at every $t$ divisible by three it slightly increases before decreasing again. 
This appears to be a direct effect of the set size parameter $s$. 

As it is computationally expensive to compute the winning ratios for larger hands, 
we also study a smaller instance with $n = 6$, $m = 2$ and $k = 4$. This leaves the number of suits 
and copies of the tile set identical to the original version, and brings the
total number of tiles approximately to half of that in the original game. 
The ratio of winning sets was calculated for all hand sizes in this configuration. 
The results are shown in Figure~\ref{fig:counts}. It shows the size of the sets
against the ratio of winning hands (logarithmic scale). We observe
the same pattern as for the original game. For small hands the ratio seems to 
decrease as the hand size increases, except for spikes at multiples of three. 
At some point, this effect diminishes and the ratio strictly increases. 
This happens from hands of size $20$ and on. 
However, the ratio is still rather low. 
Even with a hand consisting of half of all the tiles, chances of being able to 
play all tiles in one move are very low ($0.2\%$). 
It seems plausible that this percentage is even lower for the original game with $n=13$. 
This observation could be useful for some dynamics in the two-player game, e.g., for determining whether the player should play some tiles or take one from the pool.


\section{Conclusion}
\label{sec:conclusion}
We have presented a polynomial algorithm with a dynamic programming approach that can solve the Rummikub puzzle.
Next, we gave an analysis of the influence of different input parameters on the space and time complexity of both the proposed algorithm and the general problem of solving the Rummikub puzzle. 
Furthermore, we have used the algorithm to compute the percentage of hands of a given number of tiles can be played at once as part of a first attempt to understand the aspects involved in the multi-player game of Rummikub. 

In future work we plan to more elaborately address the multi-player game: although we show that the presented algorithm for solving the Rummikub puzzle can be used in the Rummikub game, the strategy aspects of the multi-player game, such as hand-making decisions, are not considered. 
Two-player games with a bound on the number of moves are generally PSPACE-complete~\cite{Hearn2007}; whether this holds for Rummikub remains an open question worth researching further. 
Last but not least, the complexity of solving the Rummikub optimization problem when the number of suits $k$ is considered as an input parameter also remains an interesting open problem: this proof would be the final piece to solving the Rummikub puzzle.

\bibliographystyle{plain}
\bibliography{rummikub}

\end{document}